# Analysis of the flooding search algorithm with OPNET

Arkadiusz Biernacki

*Abstract*—In this work we consider the popular OPNET simulator as a tool for performance evaluation of algorithms operating in peer-to-peer (P2P) networks. We created simple framework and used it to analyse the flooding search algorithm which is a popular technique for searching files in an unstructured P2P network. We investigated the influence of the number of replicas and time to live (TTL) of search queries on the algorithm performance. Preparing the simulation we did not reported the problems which are commonly encountered in P2P dedicated simulators although the size of simulated network was limited.

## I. INTRODUCTION

A large number of peer-to-peer (P2P) systems based on overlay networks have been developed in recent years. Simultaneously with the evolution of P2P systems a number of P2P overlay simulators have being developed, among them are: P2PSim, PeerSim, NeuroGrid and PlanetSim. Most of these simulators were created by various research groups for use by the P2P academic community. In spite of the simulators diversity most of them lack some important features required from this kind of software [1]. Usually the documentation is poorly written or parts of the software functionality remain undocumented. Some of the simulators, such as PlanetSim, have no means to collect statistics and those that do provide often very limited sets of variables which are available for an end user. Sometimes user wishing to change the variables for which data can be captured, will have to modify the code as required. Lack of clarity in the properties of experiments makes reproducibility of results and analysis and comparison between algorithms problematic. The full survey of P2P simulators and their suitability for simulations may be found i.e. in [1][2].

In this paper we examine suitability of OPNET, which is well known commercial discrete event simulator, as a tool for performance evaluation of P2P overlay networks. The main contribution of the paper is creation of simple framework for a simulation of unstructured P2P networks in the OPNET environment. We show practical usage of the framework analyzing the flooding search algorithm used in this type of networks.

A. Biernacki is with the Institute of Computer Science, Silesian University of Technology, Akademicka 16, 44-100 Gliwice, Poland (e-mail: arkadiusz.biernacki@polsl.pl).

## II. THEORETICAL BACKGROUND

### A. P2P overlay networks

A Peer-to-Peer (P2P) file sharing system is built as an overlay on the existing Internet infrastructure. It provides a file sharing service to a highly transient population of users (peers). Early systems, such as Napster, used a central server to store indices of participating peers. This centralized design concerns of performance bottleneck and single point of failure. To avoid such possibility, instead of maintaining a huge index in a centralized system, a decentralized system distributes all searching and locating loads across all the participating peers. Though the decentralized approach concerns the overloading and reliability issues, and it is thought to build a scalable P2P system, its success is considerably dependent on an efficient mechanism to broadcast queries (messages, packets) across a large population of peers. Reaching out to a large scope of peers is a fundamental procedure in an unstructured P2P network.

In our work we consider purely unstructured network where each peer stores a local collection of objects. Nodes generate search queries and send them through the network. Peers and objects are assumed to have unique identifiers, with object IDs used to specify the query target. Search algorithms can not in any way dictate object placement and replication in the system. They are also not allowed to alter the topology of the P2P overlay. Nodes that are directly linked in the overlay are neighbours. A node is always aware of the existence and identity of its neighbours.

To improve the search and whole system performance, the object may be replicated on certain number of hosts.

### B. Flooding algorithm

The popular technique for searching files in an unstructured network is the flooding algorithm. In the algorithm each node acts as both a transmitter and a receiver and each node tries to forward every search query to every one of its neighbours except the source node. Each search query has an unique number. A query received by a peer that has the same number as the one received previously will be discarded in order to avoid redundancy. Flooding is performed in a hop by hop fashion counted by time-to-live (TTL) counter for each query. A query starts off with its initial TTL set to specified value, which is decremented by one when it travels between two nodes. A query comes to its end either when it becomes a redundant query, when its TTL is decreased to 0 or when the data it is looking for is found. The flooding algorithm is not too effi-cient, because queries are generally broadcast indiscriminately in a whole neighbourhood using lot of network resources. As a result its search efficiency decays as the search time increases since the number of query messages increases with the size of visited peers. Consequently the algorithm faces the scalability problem when the query time increases. To mitigate those drawbacks there have been created numerous others flavours of the flooding algorithm, i.e. [3][4]. Despite the aforementioned drawbacks the algorithm advantage is that it demands very little management overhead, adapts well to the transient activity of P2P clients and takes advantage of the spontaneous replication of popular content. The flooding algorithm is used i.e. in popular Gnutella system [5]. General review of other search algorithm can be found among other in [6].

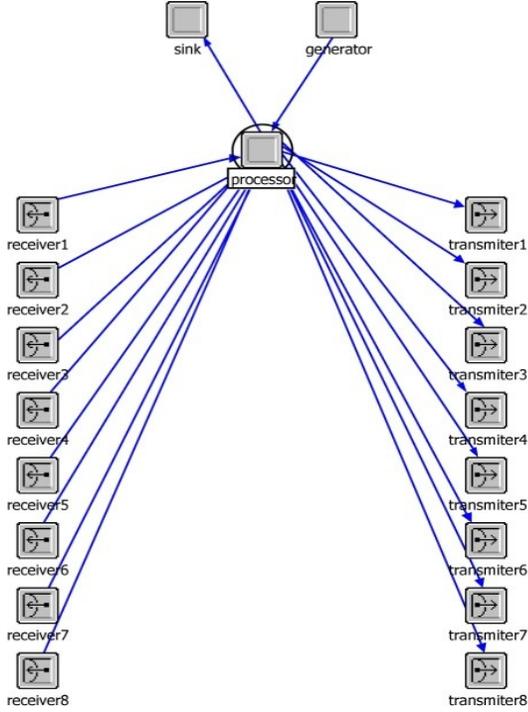

Fig. 1. Architecture of a single network node

### C. Performance metrics

To measure search performance we took metrics commonly used in the papers concerning this topic [7][8].

A search query is successful if it discovers at least one replica of the requested object. The ratio of successful to total searches made is called the *success rate*.

A single search can result in multiple discoveries (hits), which are replicas of the same object stored at distinct nodes. Number of discoveries per single search query is called the *hits per query*.

Average number of hops needed for successful search is called the *average hops number*. It is information about delay in finding an object as measured in number of hops. We did not model the actual network latency here, but rather just measured the abstract number of hops that a successful search message travelled before it replied to the originator. A global TTL parameter represents the maximum hop-distance a query can reach before it gets discarded.

Overhead of an algorithm is measured in average number of packets which the P2P network has to process per single search query. All the packets generated during a single search are called the *forwarded packets*.

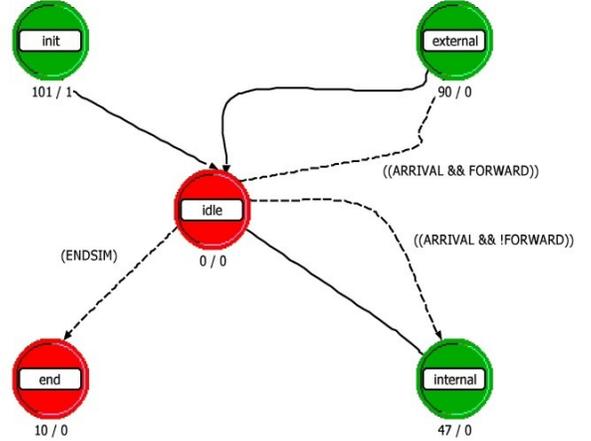

Fig. 2. State machine diagram of node's processor

### III. EXPERIMENT

#### A. Assumptions

We created an overlay peer-to-peer network consisting of 1000 nodes, the links distribution was a random variable with the uniform distribution ranging between 2 and 8. In network nodes we placed 500 distinct objects. The above assumptions were simplified; in the real the topology of P2P network like Gnutella is a two-stage power-law graph [5]. The number of nodes used in the simulation is also smaller by at least an order of magnitude compared to the real network.

Each object was assigned a unique natural number. Replication of the objects was a parameter of the simulation and for our experiment we used a set consisting of five values: 2, 8, 32, 128 and 512. When replication parameter is set to $n$ it means there are n instances of every object in the network. We used uniform strategy, replicating everything equally between nodes. Each node contained similar number of object which may be estimated by the following formula:

$SNObjCount = TObjCount \cdot RP / TNodeCount$

where: *SNObjCount* is number of objects in a single node, *TObjCount* is total number of objects, *RP* replication parameter and *TNodeCount* is total number of nodes.

In order to get better simulation performance and to simplify simulation design we did not take into account protocols stack (TCP/IP). Simulation was based only on passing search queries between networks nodes which was sufficient in the case of the flooding algorithm. The search query (packet) has 4 fields: ID, source address, searched object ID and TTL. The first two fields were not explicitly used in the communication (we did not send any acknowledgement packets concerning successful search queries), however they were used for statistical pur-poses.

In our experiment we tried to answer the following questions:
- What percent of search queries ended up with success in function of objects replication, TTL of search packets?
- How many packets were found in a single query?
- In the case of successful query what was its search time measured in search query hops?
- How many packets were generated by a single search query?

*B. Simulation architecture*

The architecture of a single network node is presented on figure 1. It consisted of a packet generator, a packets sink, transmitters and receivers through which connections were made with others nodes. Search queries were generated according to the Poisson distribution by ten of the nodes. Packets which their TTL reached 0 were forwarded to the *sink* module and destroyed. A node was connected with other nodes through bidirectional links associated to the pairs of transmitters and receivers modules. The main logic was included in the processor module based on a state machine diagram – figure 2. The *init* state in the diagram was responsible for initialization of statistic and other node parameters. Incoming packets were serviced by the *internal* or *external* states depending on whether they were newly generated or they were coming from other nodes.

IV. SIMULATION RESULTS

We gathered the experiment results on two levels – global and local. The global level concerned the averaged aggregated network statistic while on the local level we were able to examine single node behaviour. We collected four statistics from the metrics mentioned in section II C.

On the figure 3 we presented success rate per single query. As expected there is a direct relation between TTL and the success rate of a search query. The choice of proper TTL value is important in case of small number of replications, the larger the number of replicas the lower the value of TTL which is sufficient to find an object.

The hits per single query are presented on figure 4. The higher is the number of the replication the higher is the hits value except situation where the replication parameter is 512 and TTL is 8 or TTL is 6. This abnormality may be explained as the following: in the case of high number of replications the success rate is very close to one (figure 3) and the flood of packets is quickly attenuated. Search queries did not have enough time to replicate themselves because they quickly ended up as successful queries and were perished.

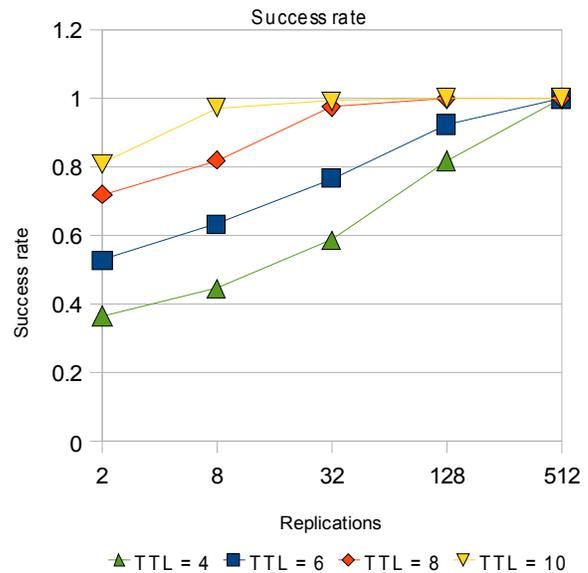

Fig. 3. Success rate per single query

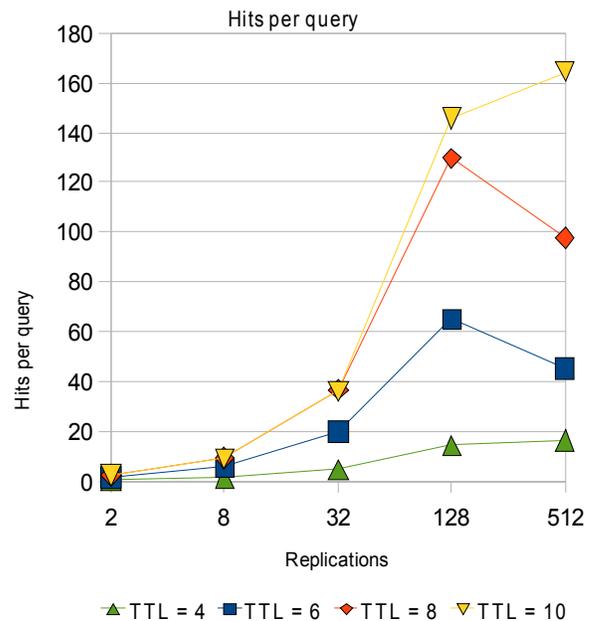

Fig. 4. Successful hits per single query

Average hops number is presented on figure 5. Except case were replications was 512 hops number are close to TTL. Such results indicate that most of the searched objects were placed relatively far from the node which invoked the search.

On the figure 6 we presented number of forwarded packets per query. The umber of forwarded packet is directly proportional to TTL and inversely proportional to the objects replications number.

The statistics related to the *success rate*, *hits per query*, *average hops number* and *forwarded packets* gathered at the global level were also gathered at the local level. Although it is possible to gather the local statistic for every node, such strategy would have slowed down the simulation speed, therefore we gathered them for only three selected nodes. The res-ults were presented on figure 7 and are with agreement with the global statistics.

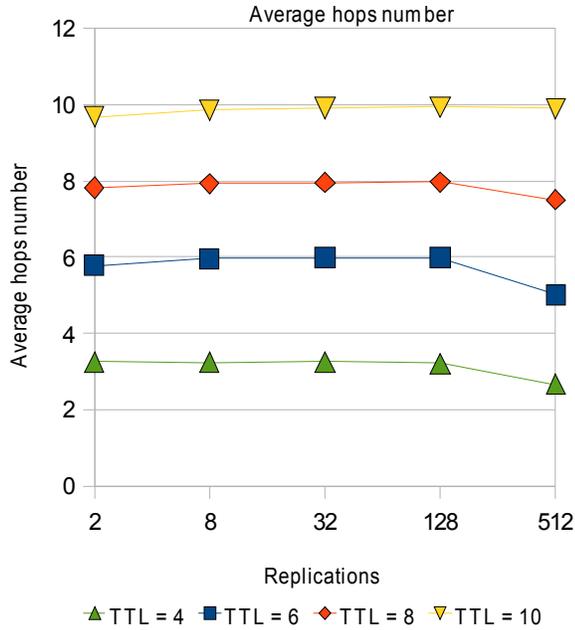

Fig. 5. Average hops number per successful query

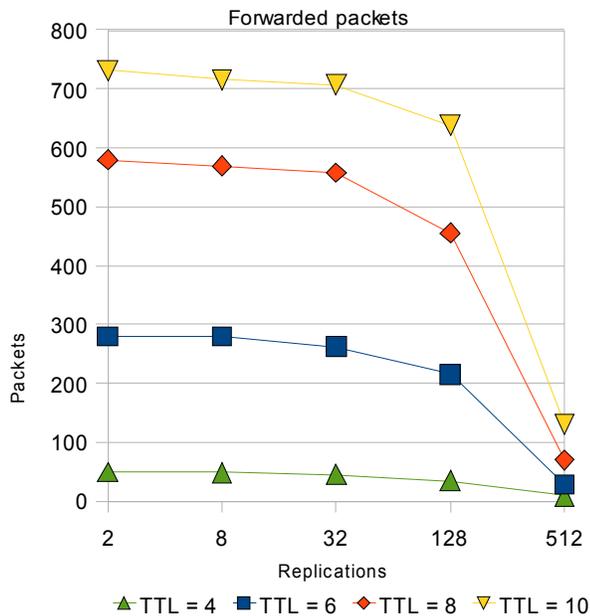

Fig. 6. Number of forwarded packets per single query

## V. CONCLUSIONS

In this work we presented analysis of flooding search algorithm, popularly used in P2P networks, using the OPNET simulator. Performing the simulation we did not reported the problems which are commonly encountered in P2P dedicated simulators. The architecture of the simulation seemed to be clear, modularized and easily scalable. Although we reported some others issues, amongst them is restricted number of nodes in the network. Tries with simulation of network consisting of more than 5000 nodes lasted much longer and sometimes OPNET stopped responding at all.

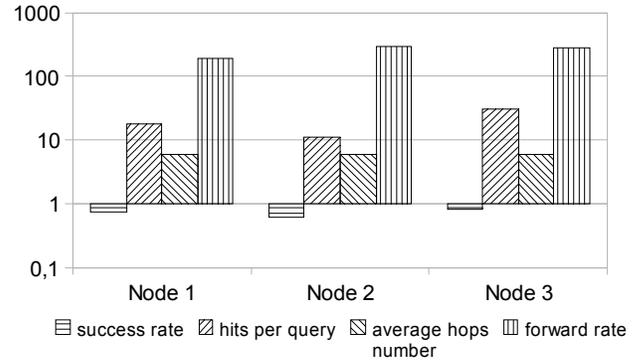

Fig. 7. Local statistic for single nodes

Despite these drawbacks we may state that the OPNET environment is suitable for a "first-look" performance evaluation of P2P algorithms. The environment may be helpful i.e. for a first assessment of the algorithm: if the results of quick analysis in the OPNET are promising than it may be simulated in a dedicated P2P simulator.

Further works will concentrate on the adoption of our framework to a simulation of more realistic models. We plan to import P2P network power-law graph topology and evaluate performance of the simulation which contains larger number of nodes. These steps should lead to analysis of more advanced search algorithms and proposals of new ones.